\documentclass[aps,prb,twocolumn,groupedaddress,amsmath]{revtex4-2}
\usepackage{graphicx}
\usepackage{amssymb}
\usepackage{threeparttable}
\begin{document}
\newcommand{\magdir}{\hat{\vn{n}}}
\newcommand{\vn}[1]{{\boldsymbol{#1}}}
\newcommand{\vht}[1]{{\boldsymbol{#1}}}
\newcommand{\polarivec}{\boldsymbol{\epsilon}}
\newcommand{\matn}[1]{{\bf{#1}}}
\newcommand{\matnht}[1]{{\boldsymbol{#1}}}
\newcommand{\bege}{\begin{equation}}
\newcommand{\ee}{\end{equation}}
\newcommand{\bal}{\begin{aligned}}
\newcommand{\defbar}{\overline}
\newcommand{\SM}{\scriptstyle}
\newcommand{\eal}{\end{aligned}}
\newcommand{\torkance}{t}
\newcommand{\udot}{\overset{.}{u}}
\newcommand{\exponential}[1]{{\exp(#1)}}
\newcommand{\phandot}[1]{\overset{\phantom{.}}{#1}}
\newcommand{\phandag}{\phantom{\dagger}}
\newcommand{\Trace}{\text{Tr}}
\newcommand{\Bxc}{\Omega}
\newcommand{\mubo}{\mu_{\rm B}^{\phantom{B}}}
\newcommand{\rmd}{{\rm d}}
\newcommand{\rme}{{\rm e}}
\newcommand{\intkspa}{\int\!\!\frac{\rmd^3 k}{(2\pi)^3}}
\newcommand{\intkspatwodim}{\int\!\!\frac{\rmd^2 k}{(2\pi)^2}}
\setcounter{secnumdepth}{2}
\title{Charge and spin photocurrents in the Rashba model}
\author{Frank Freimuth$^{1,2}$}
\email[Corresp.~author:~]{f.freimuth@fz-juelich.de}
\author{Stefan Bl\"ugel$^{1}$}
\author{Yuriy Mokrousov$^{1,2}$}
\affiliation{$^1$Peter Gr\"unberg Institut and Institute for Advanced Simulation,
Forschungszentrum J\"ulich and JARA, 52425 J\"ulich, Germany}
\affiliation{$^2$Institute of Physics, Johannes Gutenberg University Mainz, 55099 Mainz, Germany
}
\date{\today}
\begin{abstract}
In metallic noncentrosymmetric crystals
and at surfaces the response of
spin currents and charge currents to applied electric fields
contains contributions that are second order in the electric
field, which are forbidden by symmetry in
centrosymmetric systems. Thereby, photocurrents and
spin photocurrents can be generated
in inversion asymmetric metals by the application of femtosecond
laser pulses. 
We study the laser-induced charge current in the ferromagnetic
Rashba model with in-plane magnetization
and find that 
this 
\textit{magnetic photogalvanic effect} 
can be
tuned to be comparable in size
to the laser-induced photocurrents measured experimentally
in magnetic bilayer systems such as Co/Pt.
Additionally, we show that femtosecond laser pulses
excite strong spin currents in the nonmagnetic Rashba model when the
Rashba parameter is large. 
\end{abstract}

\maketitle
\section{Introduction}
The generation of in-plane charge currents by application
of ultrashort laser pulses to magnetic bilayer systems with
structural inversion asymmetry -- such as Co/Pt, Co/Ta
and Co$_{20}$Fe$_{60}$B$_{20}$/W -- is currently
attracting 
attention, because it paves the way to ultrafast electronics,
because the resulting terahertz (THz) signals can be used to develop
efficient table-top 
THz-emitters~\cite{thz_emitter_Seifert,THz_emitters_review}, and because
these charge currents
contain information about several important effects, such as 
superdiffusive spin-currents~\cite{thz_spin_current_kampfrath,spin_photocurrents_GdFeCoPt_huisman}, 
spin Hall angles, the inverse Faraday effect (IFE),
and the inverse spin-orbit torque (SOT)~\cite{femtosecond_control_electric_currents_Huisman}.
So far, two mechanisms for in-plane photocurrent generation in magnetic bilayers 
have been identified in experiments:
First, the laser pulse triggers a superdiffusive spin 
current~\cite{battiato_superdiffusive_spin_transport_ultrafast_demagnetization,PhysRevB.86.024404,Malinowski_ultrafast_demag_direct_spin_transfer,melnikov_AuFeMgO_PhysRevLett.107.076601},
which flows from the magnetic into the nonmagnetic layer and
which is converted into an in-plane electric current by the inverse
spin-Hall effect~\cite{thz_spin_current_kampfrath,thz_emitter_Seifert}. 
Second, the IFE can be used
to induce magnetization dynamics in the ferromagnetic 
layer~\cite{optical_driven_magnetization_dynamics_choi_2017}, 
which drives an in-plane
electric current due to the inverse SOT~\cite{femtosecond_control_electric_currents_Huisman,invsot}. 
Additionally, it has been shown theoretically that electric currents
are generated if the exchange splitting varies in time after laser
excitation~\cite{longitex}. Thus, the laser-induced 
photocurrents contain also information about whether ultrafast
demagnetization is dominated by an 
exchange-field collapse~\cite{Krieger_2017}
or by the excitation of transverse spin 
fluctuations~\cite{ufd_pastor}. Therefore, they can be used 
to study the nature of ultrafast demagnetization, as an alternative or 
complementary
tool to conductivity measurements~\cite{collapsed_vs_collective} 
and photoelectron spectroscopy~\cite{Eiche1602094}.

Circularly polarized light induces an electric current even in 
noncentrosymmetric nonmagnetic semiconductors, which is known
as the circular photogalvanic effect~\cite{spin_photocurrents_quantum_wells,direct_optical_detection_weyl_fermion_chirality_topological_semimetal,control_ti_photocurrents_light_polarization}. 
The question therefore arises
whether in noncentrosymmetric magnetic metals there exists
an effect similar to the circular photogalvanic effect
and whether such an effect contributes to the
laser-induced charge currents in magnetic bilayer systems~\cite{PhysRevMaterials.3.084415}.
Effects from the interfacial spin-orbit interaction (SOI) in magnetic bilayer systems can
be studied based on the Rashba model~\cite{rashba_review}.
In the nonmagnetic Rashba model light 
can induce out-of-plane charge currents only and no in-plane
charge currents due to symmetry. 
However, the magnetization vector in magnetic bilayers lowers the symmetry
and one may thus expect additional electric currents perpendicular
to the light wave vector, i.e., in-plane charge currents when the
light wave vector is perpendicular to the bilayer interface and when
the magnetization is in-plane.   
This effect can be considered as the \textit{magnetic photogalvanic effect}.

Also pure spin currents can be excited by 
light
in noncentrosymmetric nonmagnetic 
semiconductors~\cite{pure_spin_current_one_photon_absorption,injection_ballistic_pure_spin_currents,zero_bias_spin_separation},
in graphene~\cite{photoinduced_pure_spin_current_graphene} 
deposited on a substrate or subject to an
external out-of-plane electric field,
and in organic-inorganic 
halide CH$_3$NH$_3$PbI$_3$~\cite{organic_inorganic_perovskite_photovoltaics_PhysRevB.93.155432},
which is an important step towards ultrafast spintronics.
Since a very strong Rashba effect has been found 
in Bi/Ag(111) surface alloys~\cite{giant_spin_splitting_surface_alloying},
one may expect very efficient generation of
spin photocurrents in this metallic surface, which would make
Bi/Ag(111) attractive for ultrafast metallic spintronics applications.

In this work we study the laser-induced in-plane
charge currents in the ferromagnetic Rashba model in order to
find out how large photocurrents can be that
are generated directly by the interfacial Rashba SOI in
magnetic bilayer systems without
involving the generation of superdiffusive spin-currents or the
excitation of magnetization dynamics through the
IFE. Thereby we extend the list
of suggested mechanisms for the generation of in-plane photocurrents
by light in magnetic bilayer systems.
In view of the discovery of more and more 
nonmagnetic systems with
a giant Rashba effect~\cite{giant_ambipolar_rashba_BiTeI,giant_rashba_spin_splitting_BiTeI,strongly_soc_coupled_2deg_PhysRevLett.110.107204,giant_spin_splitting_surface_alloying,large_tunable_rashba_spin_splitting_Bi2Se3,giant_rashba_PbSnTe}
we investigate also the laser-induced pure spin current in the
nonmagnetic Rashba model for large SOI strength,
in order to show that very strong spin currents can be generated optically
in such materials.

This paper is organized as follows.
In Sec.~\ref{sec_formalism} we describe the formalism
that we use to compute the laser-induced charge currents
and spin currents, which is based on the Keldysh
nonequilibrium formalism.
In Sec.~\ref{sec_symmetry} we discuss based on
symmetry arguments which
components of the laser-induced charge currents
and spin currents can exist and
which components are zero in the Rashba model.
In Sec.~\ref{sec_results_charge_current} 
we present numerical results for the laser-induced charge current
in the ferromagnetic Rashba model and
in Sec.~\ref{sec_results_spin_current} 
we discuss the numerical results for the laser-induced
spin current in the nonmagnetic Rashba model.
This paper ends with a summary in Sec.~\ref{sec_summary}.

\section{Formalism}
\label{sec_formalism}
In experiments femtosecond
laser pulses are used very often instead of continuous laser beams,
because thereby much larger electric field strengths can
be applied. Additionally, a novel type of THz emitter
uses excitation by femtosecond laser 
pulses~\cite{thz_emitter_Seifert,THz_emitters_review}. 
We assume that the response to laser pulses
can be modelled by considering the time-dependent
intensity $I(t)$ of the laser pulse and by assuming that
the response at time $t$ agrees with the hypothetical
response to a continuous laser beam with constant 
intensity $I$ given by $I=I(t)$.  
This approximation is justified for the parameters used in this work, because
50-fs pulses of light wavelength 800 nm 
correspond to roughly 20 oscillations of the electric field vector.
Moreover, this approximation has predicted the amplitudes of the effects
satisfactorily~\cite{lasintor}.

\subsection{Laser-induced charge current}
The response that arises at the second order
in the perturbing electric field of a 
continuous laser beam with frequency $\omega$
contains a dc contribution and an ac contribution with 
frequency $2\omega$. Here, we are only interested
in the dc contribution. 
The dc electric current response to a continuous laser beam
appears as a THz electric current pulse when femtosecond laser
pulses are used instead of a continuous laser beam.
Therefore, in the following we discuss the expressions to
compute the dc electric current driven by a continuous
laser beam with light frequency $\omega$.

To derive expressions suitable to
describe the laser-induced electric current
one can proceed in close analogy 
to the case of torques that arise at the second order 
in the perturbing electric field of the laser,
which were discussed in detail in Ref.~\cite{lasintor}.
We do not present the detailed derivation here but only the final expression.
The dc electric 
current density 
that arises at second order in the electric field of a continuous
laser
beam of frequency $\omega$ can be written as
\bege\label{eq_define_current_chi}
J_{\alpha}=\frac{a^{2}_{0} e I}{\hbar c}
\left(
\frac{\mathcal{E}_{\rm H}}{\hbar\omega}
\right)^2
{\rm Im}
\sum_{\beta\gamma}
\epsilon_{\beta}^{\phantom{i}}
\epsilon_{\gamma}^*
\varphi_{\alpha\beta\gamma}^{\phantom{i}},
\ee
where $\varphi_{\alpha\beta\gamma}^{\phantom{i}}=\chi_{\beta\gamma}^{v_{\alpha}}$ 
and the response of a general observable $\mathcal{O}$ to monochromatic
light of frequency $\omega$ is given by
\bege\label{eq_chi_noabrev}
\begin{aligned}
\chi^{\mathcal{O}}_{\beta\gamma}=&\frac{2}{a_0\mathcal{E}_{\rm H}}
\intkspatwodim
\int \rmd \mathcal{E}\,
{\rm Tr}\Big[\\
&f(\mathcal{E})
\mathcal{O}
G^{\rm R}_{\vn{k}}(\mathcal{E})
v_{\beta}
G^{\rm R}_{\vn{k}}(\mathcal{E}-\hbar\omega)
v_{\gamma}
G^{\rm R}_{\vn{k}}(\mathcal{E})
\\
-&
f(\mathcal{E})
\mathcal{O}
G^{\rm R}_{\vn{k}}(\mathcal{E})
v_{\beta}
G^{\rm R}_{\vn{k}}(\mathcal{E}-\hbar\omega)
v_{\gamma}
G^{\rm A}_{\vn{k}}(\mathcal{E})\\
+&f(\mathcal{E})
\mathcal{O}
G^{\rm R}_{\vn{k}}(\mathcal{E})
v_{\gamma}
G^{\rm R}_{\vn{k}}(\mathcal{E}+\hbar\omega)
v_{\beta}
G^{\rm R}_{\vn{k}}(\mathcal{E})
\\
-&f(\mathcal{E})
\mathcal{O}
G^{\rm R}_{\vn{k}}(\mathcal{E})
v_{\gamma}
G^{\rm R}_{\vn{k}}(\mathcal{E}+\hbar\omega)
v_{\beta}
G^{\rm A}_{\vn{k}}(\mathcal{E})
\\
+&
f(\mathcal{E}-\hbar\omega)
\mathcal{O}
G^{\rm R}_{\vn{k}}(\mathcal{E})
v_{\beta}
G^{\rm R}_{\vn{k}}(\mathcal{E}-\hbar\omega)
v_{\gamma}
G^{\rm A}_{\vn{k}}(\mathcal{E})\\
+&
f(\mathcal{E}+\hbar\omega)
\mathcal{O}
G^{\rm R}_{\vn{k}}(\mathcal{E})
v_{\gamma}
G^{\rm R}_{\vn{k}}(\mathcal{E}+\hbar\omega)
v_{\beta}
G^{\rm A}_{\vn{k}}(\mathcal{E})
\Big].\\
\end{aligned}
\ee
Here, $a_{0}^{\phantom{i}}=4\pi\epsilon_{0}^{\phantom{i}}\hbar^2/(me^2)$ is Bohr's
radius, $I$ is the
intensity of light, $c$ is the velocity of
light, $\mathcal{E}_{\rm H}=e^2/(4\pi\epsilon_{0}^{\phantom{i}} a_{0}^{\phantom{i}})$ is the Hartree
energy and $f(\mathcal{E})$ is the Fermi distribution
function.
$v_{\beta}$ is the $\beta$th component of the velocity
operator, $e$ is the elementary positive charge, $\mathcal{E}_{\rm F}^{\phantom{i}}$ is
the Fermi energy,
\bege \label{eq_define_green_analy}
G^{\rm R}_{\vn{k}}(\mathcal{E})=
\hbar
\sum_{n}\frac{|\vn{k}n\rangle\langle\vn{k}n|}{\mathcal{E}-\mathcal{E}_{\vn{k}n}+i\Gamma},
\ee
is the retarded Green function and
$G^{\rm A}_{\vn{k}}(\mathcal{E})=[G^{\rm R}_{\vn{k}}(\mathcal{E})]^{\dagger}$
is the advanced Green function. The energy of the
state $|\vn{k}n\rangle$ of an electron in band $n$ at $k$-point $\vn{k}$
is $\mathcal{E}_{\vn{k}n}$. The parameter $\Gamma$ describes the
lifetime broadening of the electronic states.
$\epsilon_{\beta}^{\phantom{i}}$ is the $\beta$th component of the polarization
vector of the light.
Circularly polarized light with light wave vector along the $z$ direction
is described by $\vn{\epsilon}=(1,\lambda i,0)/\sqrt{2}$, 
where $\lambda=\pm 1$ controls the light helicity.
In order to obtain the 
coefficients $\varphi_{\alpha\beta\gamma}=\chi_{\beta\gamma}^{v_{\alpha}}$
in Eq.~\eqref{eq_define_current_chi}
one only needs to substitute $\mathcal{O}$ in Eq.~\eqref{eq_chi_noabrev} 
by $v_{\alpha}$, i.e., by the $\alpha$th component 
of the velocity operator.
Note that while the Green's function is chosen to be band-diagonal
in Eq.~\eqref{eq_define_green_analy} the velocity operator 
and $\mathcal{O}$ are not band-diagonal and induce
interband transitions in Eq.~\eqref{eq_chi_noabrev}.

\subsection{Laser-induced spin current}
Similarly, the dc spin-current density that arises in second order 
response to the electric field of the laser 
can be written as
\bege\label{eq_define_spin_current_chi}
J_{\alpha}^{s}=-\frac{a^2_{0} I}{4 c}
\left(
\frac{\mathcal{E}_{\rm H}}{\hbar\omega}
\right)^2
{\rm Im}
\sum_{jk}
\epsilon_{\beta}^{\phantom{i}}
\epsilon_{\gamma}^*
\phi_{\alpha\beta\gamma}^{s},
\ee
where $\phi_{\alpha\beta\gamma}^{s}=\chi_{\beta\gamma}^{\{ v_{\alpha},\sigma_{s}\} }$,
i.e., $\phi_{\alpha\beta\gamma}^{s}$ is obtained 
from $\chi_{\beta\gamma}^{\mathcal{O}}$ in Eq.~\eqref{eq_chi_noabrev}
by the substitution $\mathcal{O}=\{ v_{\alpha},\sigma_{s}\}$.
Here, $\sigma_{s}$ ($s=x,y,z$) are the Pauli spin-matrices and
$J_{\alpha}^{s}$ is the component 
of the spin-current density
where
the spin of the carriers 
is oriented in $s$ direction and the carriers move along the $\alpha$ direction. 
\subsection{Rashba model}
\label{sec_rashba}
We investigate the laser-induced charge current and spin current
in the Rashba model (see Ref.~\cite{rashba_review} for a recent review
on the Rashba model). 
The Rashba model with an additional
exchange splitting is given by
\bege\label{eq_rashba_model}
H^{\rm R}=\frac{-\hbar^2}{2m_e}
\Delta-i
\alpha^{\rm R} (\vn{\nabla}\times\hat{\vn{e}}_{z})\cdot\vn{\sigma}+
\frac{\Delta V}{2}
\vn{\sigma}
\cdot
\magdir(\vn{r})
,
\ee
where the first, second and third terms on the right-hand side
describe the kinetic energy,
the Rashba SOI
and the exchange interaction, respectively.
By modelling the experimentally measured Dzyaloshinskii-Moriya interaction the Rashba parameter 
in Co/Pt bilayers was estimated to be
$\alpha^{\rm R}=0.095$eV\AA~\cite{sot_dmi_stiles}.
The same order of magnitude of $\alpha^{\rm R}$ was estimated for
Ni$_{80}$Fe$_{20}$/Pt~\cite{PhysRevB.97.094407}
Substantially larger values of $\alpha^{\rm R}$ have been reported for
Bi/Ag(111) surface 
alloys ($\alpha^{\rm R}=3.05$eV\AA~\cite{giant_spin_splitting_surface_alloying}),
for BiTeI ($\alpha^{\rm R}=3.85$eV\AA~\cite{giant_rashba_spin_splitting_BiTeI}),
and for Pb$_{1-x}$Sn$_{x}$Te 
($\alpha^{\rm R}=3.8$eV\AA~\cite{giant_rashba_PbSnTe}).

In the Rashba model the velocity 
operator $\vn{v}_{\vn{k}}^{\phantom{k}}=e^{-i\vn{k}\cdot\vn{r}}\vn{v}e^{i\vn{k}\cdot\vn{r}}$ 
is given by
\bege
\vn{v}_{\vn{k}}^{\phantom{k}}=\frac{1}{\hbar}\frac{\partial H^{\rm R}}{\partial\vn{k}}=
\frac{\hbar}{m}\vn{k}+\alpha^{\rm R}\hat{\vn{e}}_{z}\times\vn{\sigma}.
\ee
Due to the term $\alpha^{\rm R}\hat{\vn{e}}_{z}\times\vn{\sigma}$ the
velocity operator $\vn{v}_{\vn{k}}^{\phantom{k}}$ does not commute
with the Hamiltonian $H^{\rm R}$. Consequently, the eigenstates of the
Hamiltonian $H^{\rm R}$ are not simultaneously eigenstates of the 
velocity operator $\vn{v}_{\vn{k}}^{\phantom{k}}$.
When the Green's function is chosen to be band-diagonal like 
in Eq.~\eqref{eq_define_green_analy} the velocity operator is
not band-diagonal and induces interband transitions 
in Eq.~\eqref{eq_chi_noabrev}.
\section{Symmetry properties}
\label{sec_symmetry}
The response of the charge current to the second-order perturbation
by an applied electric field is described by a polar tensor of third rank.
Therefore, it is nonzero only in noncentrosymmetric crystals.
Similarly, the response of the spin-current to the second order
perturbation by an applied electric field, which is described by
an axial tensor of fourth rank, 
is nonzero only in noncentrosymmetric crystals~\cite{Ivchenko2008}.
In the following we discuss which components of the laser-induced
charge current and of the laser-induced spin current are allowed by
symmetry
in the Rashba model.
\subsection{Laser-induced charge current}
\label{sec_symmetry_charge_current}
Circularly polarized light with wave vector parallel to the $z$
direction
does not induce in-plane charge currents in the
nonmagnetic Rashba model 
due to the rotational symmetry around the $z$ axis: $c_2$ rotation
flips both $J_x$ and $J_y$ but does not change the 
helicity $\lambda$ (see Table~\ref{tab_lmj_symmetry}).
Similarly, linearly polarized light does not induce in-plane
charge currents in this case.

\begin{threeparttable}
\caption{Effect of $c_2$ rotation around the $z$ axis,
effect of mirror reflection $\mathcal{M}_{zx}$ at the $zx$
  plane, and effect of
mirror reflection $\mathcal{M}_{yz}$ at the $yz$ plane on
light helicity $\lambda$,
magnetization $\vn{M}$,
current density $\vn{J}$, 
and spin current density $J_{\alpha}^{s}$.
The magnetization
$\vn{M}$ transforms like an axial vector,
while the current  density $\vn{J}$ transforms like a polar vector.}
\label{tab_lmj_symmetry}
\begin{ruledtabular}
\begin{tabular}{c|c|c|c|c|c|c|c|c|c|c}
 &$\lambda$
&$M_{y}$ &$J_{x}$ &$J_{y}$ &$J_{x}^{y}$ &$J_{y}^{x}$
 &$J_{x}^{x}$ &$J_{y}^{y}$ &$J_{x}^{z}$ &$J_{y}^{z}$\\
\hline
$c_2$ &$\lambda$
 &$-M_{y}$ &$-J_{x}$ &$-J_{y}$ &$J_{x}^{y}$ &$J_{y}^{x}$
 &$J_{x}^{x}$ &$J_{y}^{y}$&$-J_{x}^{z}$ &$-J_{y}^{z}$\\
\hline
$\mathcal{M}_{zx}$&$-\lambda$
 &$M_{y}$ &$J_{x}$ &$-J_{y}$ &$J_{x}^{y}$ &$J_{y}^{x}$
 &$-J_{x}^{x}$ &$-J_{y}^{y}$&$-J_{x}^{z}$ &$J_{y}^{z}$\\
\hline
$\mathcal{M}_{yz}$&$-\lambda$
 &$-M_{y}$ &$-J_{x}$ &$J_{y}$ &$J_{x}^{y}$ &$J_{y}^{x}$
 &$-J_{x}^{x}$ &$-J_{y}^{y}$ &$J_{x}^{z}$ &-$J_{y}^{z}$\\
\end{tabular}
\end{ruledtabular}
\end{threeparttable}

In the ferromagnetic Rashba model with magnetization along $y$
light polarized linearly along $x$ or $y$ induces a
current density $J_x$ that is odd in magnetization, because $\mathcal{M}_{yz}$
(or alternatively $c_2$)
flips the magnetization (axial vector) and $J_x$.
$J_y$ is not allowed by symmetry due to $\mathcal{M}_{zx}$, which 
leaves the magnetization invariant but flips
$J_{y}$.

In the ferromagnetic Rashba model with magnetization along $y$
circularly polarized light induces $J_y$, which is odd in the helicity
of light and odd in magnetization, because 
$\mathcal{M}_ {yz}$ 
flips the light helicity and the magnetization but preserves $J_y$, 
while $\mathcal{M}_{zx}$
preserves the magnetization, but flips the light helicity and $J_y$.
In this case symmetry allows also a nonzero $J_x$, which is even in the
helicity of light and odd in magnetization, because
$\mathcal{M}_{zx}$
flips the light helicity and preserves $J_x$ and
the magnetization, while 
$\mathcal{M}_ {yz}$
flips the light helicity, the magnetization and $J_{x}$.

These symmetry properties of the magnetic photogalvanic effect
in the ferromagnetic Rashba model
are summarized in Table~\ref{tab_lasincuc}.

\begin{threeparttable}
\caption{Symmetry properties of the magnetic photogalvanic effect 
in the ferromagnetic Rashba model with
magnetization parallel to the $y$ axis. $\emptyset$ means no effect. $M_y$ means
odd in magnetization, i.e., the effect changes sign when the magnetization is
antiparallel to the $y$ axis. $\lambda M_y$ means odd in the light helicity and odd
in the magnetization. $|\lambda| M_y$ means even in the light helicity and odd in
the magnetization.}
\label{tab_lasincuc}
\begin{ruledtabular}
\begin{tabular}{c|c|c|}
&circularly polarized
&linearly polarized ($\vn{\epsilon} || x$ or $\vn{\epsilon} || y $)
\\
\hline
$J_x$ & $|\lambda| M_y$ &$M_y$ \\
\hline
$J_y$ &$\lambda M_y$ &$\emptyset$ \\
\end{tabular}
\end{ruledtabular}
\end{threeparttable}

\subsection{Laser-induced spin current}
\label{sec_symmetry_spicu}
\subsubsection{Nonmagnetic Rashba model}
We first discuss the symmetry properties of laser-induced spin
currents in the nonmagnetic Rashba model.

\textit{For light polarized linearly along $x$} the spin-current density $J_x^y$
is allowed by symmetry:  $\mathcal{M}_{zx}$   
 does not flip $J_x^y$.
$\mathcal{M}_{yz}$   
does not flip $J_x^y$ either, because it flips
both the velocity of the carriers (polar vector) and their spin (axial vector).
In this case also $J_y^x$ is allowed by symmetry:
$\mathcal{M}_{yz}$   
does not flip $J_y^x$.
$\mathcal{M}_{zx}$   
does not flip $J_y^x$ either, because it flips
both the velocity of the carriers and their spin.
However, $J_x^x$ is forbidden by symmetry in this case, 
because 
$\mathcal{M}_{zx}$   
flips only the spin of the
carriers and not their velocity and therefore it flips $J_x^x$.
Similarly, $J_y^y$ is forbidden by symmetry in this case, 
because 
$\mathcal{M}_{yz}$   
flips $J_y^y$.
Finally, also $J_x^z$ and $J_y^z$ are forbidden by
symmetry in this case, because 
$\mathcal{M}_{zx}$   
flips $J_x^z$,
and 
$\mathcal{M}_{yz}$   
flips $J_y^z$.

\textit{For circularly polarized light} $J_x^y$ and $J_y^x$
are allowed by symmetry, if they are even in the
helicity of light, because
both 
$\mathcal{M}_{zx}$   
and 
$\mathcal{M}_{yz}$   
flip the helicity of the light. 
Since 
$\mathcal{M}_{zx}$   
flips $J_x^z$
but 
$\mathcal{M}_{yz}$   
does not, $J_x^z$ is forbidden by
symmetry. Similarly, $J_y^z$ is forbidden by
symmetry.
Both 
$\mathcal{M}_{zx}$   
and 
$\mathcal{M}_{yz}$   
flip $J_x^x$ and $J_y^y$. Therefore, $J_x^x$ and $J_y^y$
are allowed by symmetry, if they are odd in the helicity of light.

The symmetry properties of the laser-induced spin current
in the nonmagnetic Rashba model are summarized in
Table~\ref{tab_lasincuspi_magy} ($\{\}$-brackets in the Table).

\subsubsection{Ferromagnetic Rashba model}
Next, we discuss the symmetry properties of the
laser-induced spin currents in the
ferromagnetic Rashba model with magnetization along $y$.

\textit{Linearly polarized light with polarization along $x$:}
$\mathcal{M}_{zx}$   
does not flip $J_x^y$ and preserves 
the magnetization.
$\mathcal{M}_{yz}$   
flips the magnetization, 
but it does not flip $J_x^y$ because it flips
both the carrier velocity and the spin. 
Thus,
$J_x^y$ is even in magnetization.
$\mathcal{M}_{zx}$   
does not flip $J_y^x$, because it flips
both the carrier velocity and the spin. It also preserves the magnetization.
$\mathcal{M}_{yz}$   
does not flip $J_y^x$, but it flips the magnetization.
Thus, $J_y^x$ is even in magnetization.
$J_x^x$ is forbidden by symmetry, because 
$\mathcal{M}_{zx}$   
flips $J_x^x$
but preserves the magnetization.
Similarly, $J_y^y$ is forbidden by symmetry, 
because 
$\mathcal{M}_{zx}$   
flips $J_y^y$ and
preserves the magnetization.
Also $J_x^z$ is forbidden by symmetry,
because 
$\mathcal{M}_{zx}$   
flips $J_x^z$ and preserves 
the magnetization.
$\mathcal{M}_{zx}$   
preserves $J_y^z$ and the magnetization, while
the $yz$ mirror plane 
flips $J_y^z$ and the magnetization. Consequently,
 $J_y^z$ is allowed by symmetry and it is odd in the magnetization.

\textit{Circularly polarized light:}
Both
$\mathcal{M}_{zx}$   
and 
$\mathcal{M}_{yz}$   
flip the helicity of the light. Thus, $J_x^y$ and
$J_y^x$ are allowed by symmetry, if they are even in the
helicity of the light. Since 
$\mathcal{M}_{yz}$   
flips the magnetization
while 
$\mathcal{M}_{zx}$   
preserves it, $J_x^y$ and $J_y^x$
are even in the magnetization.
$\mathcal{M}_{zx}$   
flips $J_x^z$, flips the helicity and preserves 
the magnetization.
$\mathcal{M}_{yz}$   
preserves $J_x^z$, flips the helicity and flips 
the magnetization.
The combination of 
$\mathcal{M}_{zx}$   
and 
$\mathcal{M}_{yz}$   
flips $J_x^z$, flips the magnetization and preserves the helicity.
Thus, $J_x^z$ is odd in the magnetization and odd in the helicity.
$\mathcal{M}_{zx}$   
preserves $J_y^z$, flips the helicity and 
preserves the magnetization.
$\mathcal{M}_{yz}$   
flips $J_y^z$, flips the helicity and 
flips the magnetization.
The combination of 
$\mathcal{M}_{zx}$   
and 
$\mathcal{M}_{yz}$   
flips $J_y^z$, flips the magnetization and preserves the helicity.
Thus, $J_y^z$ is odd in the magnetization and even in the helicity.
Both 
$\mathcal{M}_{zx}$   
and 
$\mathcal{M}_{yz}$   
flip $J_x^x$ and $J_y^y$. Therefore, $J_x^x$ and $J_y^y$
are odd in the helicity of light.
Since 
$\mathcal{M}_{yz}$   
flips the magnetization
while 
$\mathcal{M}_{zx}$   
preserves it, $J_x^x$ and $J_y^y$
are even in the magnetization.

The symmetry properties of the laser-induced spin current
in the magnetic Rashba model with magnetization in $y$ direction
are summarized in
Table~\ref{tab_lasincuspi_magy} ($[]$-brackets in the Table).

\begin{threeparttable}
\caption{Symmetry properties of the 
laser-induced spin-current density in the 
nonmagnetic Rashba model (shown in $\{\}$-brackets)
and of the ferromagnetic Rashba model 
with magnetization parallel to
the $y$ axis (shown in $[]$-brackets).
$\emptyset$ indicates that there is no effect, while $\checkmark$ signals
that there is one. 
$M_y$ means that the effect is odd in the magnetization, i.e., it
changes sign when the magnetization is antiparallel to the $y$ axis.
$|M_y|$ signals that the effect is even in the magnetization.
Effects that are even and odd in the helicity of light are
indicated by $|\lambda|$ and 
 $\lambda$, respectively. }
\label{tab_lasincuspi_magy}
\begin{ruledtabular}
\begin{tabular}{c|c|c|}
&circularly polarized
&linearly polarized ($\vn{\epsilon}||x$ or $\vn{\epsilon}||y$)
\\
\hline
$J_x^x$ &$\{\lambda\}$,$[\lambda|M_y|]$ &$\{\emptyset\}$,$[\emptyset]$ \\
\hline
$J_x^y$ &$\{|\lambda|\}$,$[|\lambda||M_y|]$ &$\{\checkmark\}$,$[|M_y|]$ \\
\hline
$J_x^z$ &$\{\emptyset\}$,$[\lambda M_y$] &$\{\emptyset\}$,$[\emptyset]$ \\
\hline
$J_y^x$ &$\{|\lambda|\}$,$[|\lambda||M_y|]$ & $\{\checkmark\}$,$[|M_y|]$ \\
\hline
$J_y^y$ &$\{\lambda\}$,$[\lambda|M_y|]$ &$\{\emptyset\}$,$[\emptyset]$ \\
\hline
$J_y^z$ &$\{\emptyset\}$,$[M_y|\lambda|]$ &$\{\emptyset\}$,$[M_y]$ \\
\end{tabular}
\end{ruledtabular}
\end{threeparttable}

\section{Results}
In the following we present results for the
charge-current density
and for the spin-current density
induced by a continuous laser beam,
which we calculate from Eq.~\eqref{eq_define_current_chi} and
Eq.~\eqref{eq_define_spin_current_chi}, respectively.
In all results presented below, we assume that the
intensity is given by $I=10$GW/cm$^2$
and the photon energy is set to $\hbar\omega$=1.55~eV.
This photon energy corresponds to 800~nm, which is often
used in experiments on Co/Pt 
bilayers~\cite{femtosecond_control_electric_currents_Huisman,thz_spin_current_kampfrath,thz_emitter_Seifert,PhysRevMaterials.3.084415}. 
The intensity $I=10$GW/cm$^2$ is of the same order of magnitude as the
one used in these experiments.

While the Rashba model has been used successfully to model spintronic
properties qualitatively~\cite{rashba_review,sot_dmi_stiles}, precise
quantitative predictions are usually not possible due to the
simplicity of this model. 
By construction, the Rashba model misses completely the
contribution that arises from the superdiffusive
spin-currents~\cite{thz_spin_current_kampfrath}.
Also, the contribution originating from laser-induced
magnetization dynamics 
is not considered below~\cite{femtosecond_control_electric_currents_Huisman}. 
Therefore, in the following we aim only at
qualitative
results and order of magnitude estimates of laser-induced
photocurrents in Co/Pt bilayers that arise from photogalvanic effects.
Additionally, we study a photogalvanic effect of spin-currents in the non-magnetic
Rashba model.

\subsection{Laser-induced charge currents}
\label{sec_results_charge_current}
\begin{figure}
\includegraphics[width=\linewidth]{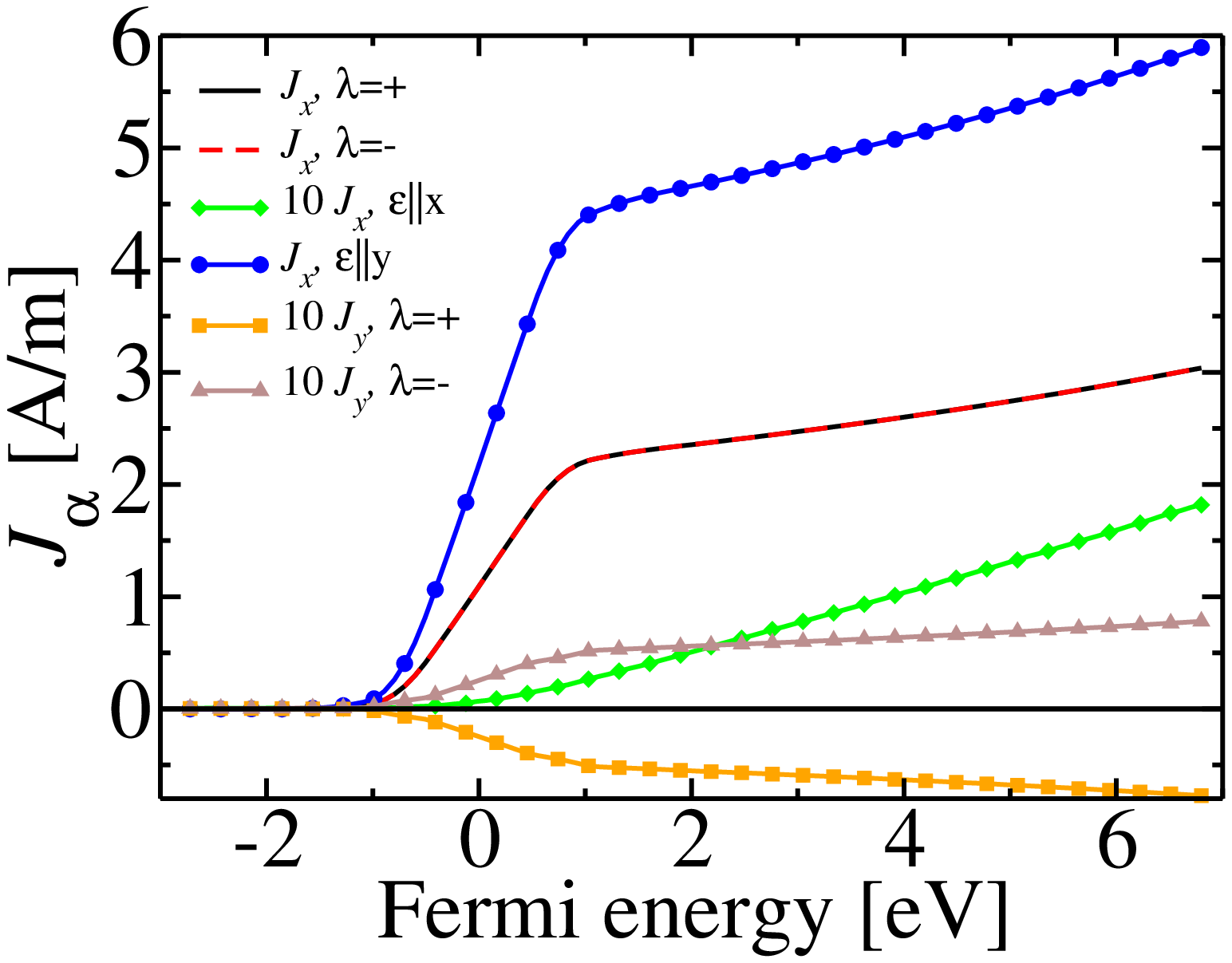}
\caption{\label{nonlinsurf_magy_vs_fermi_mimicopt_gamma25meV}
Laser-induced charge-current density $J_{\alpha}$
vs.\ Fermi energy $\mathcal{E}_{\rm F}$ for $\hat{\vn{n}}$
in $y$ direction, $\alpha^{\rm R}=0.1$eV\AA\,, $\Delta V=1$~eV,
and $\Gamma=25$meV. Some curves have been multiplied by 
the factor 10 for better visibility, as indicated in the legend.
}
\end{figure}

The laser-induced charge-current density as a function of 
Fermi energy $\mathcal{E}_{\rm F}$ is
shown in Fig.~\ref{nonlinsurf_magy_vs_fermi_mimicopt_gamma25meV}
for the parameters $\alpha^{\rm R}=0.1$eV\AA, $\Gamma=25$meV and
$\Delta V=1$~eV when $\hat{\vn{n}}$ points in $y$ direction.
As discussed in section~\ref{sec_rashba} $\alpha^{\rm R}=0.1$eV\AA\, 
is a suitable choice to model magnetic bilayer systems such as Co/Pt.
Previously, we found that the broadening of $\Gamma=25$meV is suitable
to reproduce the experimentally measured SOTs in 
\textit{ab-initio} calculations of Co/Pt bilayers~\cite{ibcsoit}
and therefore we use it here as well. 
The laser-induced current density $J_x$ is even in the helicity $\lambda$ and
for circularly polarized light
it is much larger
than for linearly polarized light with polarization along $x$. However,
for linearly polarized light with polarization along $y$ the laser-induced
current density $J_x$ is larger than the one for circularly polarized light by around
a factor of 2.
The laser-induced current density $J_y$ is odd in the helicity $\lambda$
and therefore it vanishes for linearly polarized light (not shown in the figure). 
The finding that $J_x$ is even in $\lambda$ 
while $J_y$ is odd in $\lambda$ is consistent with the symmetry analysis in 
section~\ref{sec_symmetry_charge_current}.

$J_{\alpha}$ in Fig.~1 starts to be nonzero roughly at -1eV and it
displays kinks roughly at 1eV.
In Ref.~\cite{lasintor} in Eq.~(A12) we have shown that the Fermi surface
contribution can be recast as a sum of two terms that 
contain the factors $[f(\mathcal{E}-\hbar\omega)-f(\mathcal{E})]$ 
and $f[(\mathcal{E}+\hbar\omega)-f(\mathcal{E})]$, respectively.
The first factor is nonzero 
for $\mathcal{E}_{\rm F}<\mathcal{E}<\mathcal{E}_{\rm F}+\hbar\omega$,
while the second factor is nonzero for
$\mathcal{E}_{\rm F}-\hbar\omega<\mathcal{E}<\mathcal{E}_{\rm F}$.
$J_{\alpha}$ in
Fig.~\ref{nonlinsurf_magy_vs_fermi_mimicopt_gamma25meV} 
starts to be nonzero roughly when
$\mathcal{E}_{\rm F}+\hbar\omega>$0.5eV, i.e., roughly at
$\mathcal{E}_{\rm F}>$-1eV, 
because this is when the
energy $\mathcal{E}$ can reach up to the bottom of the second band at 0.5eV
due to the first factor. The second factor is most efficient when
$\mathcal{E}_{\rm F}-\hbar\omega$
reaches the bottom of the first band, which is at -0.5eV. This happens
roughly at  $\mathcal{E}_{\rm F}>$1eV. At this energy there are kinks
in the $J_{\alpha}$ in Fig.~1.

$J_y$ increases with Fermi energy and in the range shown 
in Fig.~\ref{nonlinsurf_magy_vs_fermi_mimicopt_gamma25meV} it
is maximally 0.78mA/cm for $\mathcal{E}_{\rm F}=$6.8eV.
Experimentally, the amplitude of the current density $J_y$ 
has been estimated to be 5mA/cm when a 50fs laser pulse with fluence 1mJcm$^{-2}$
is applied to Co/Pt bilayers~\cite{femtosecond_control_electric_currents_Huisman}.
Assuming a Gaussian-shaped laser pulse
we estimate the peak intensity of the pulse
to be $I\approx 2\sqrt{{\rm ln}(2)/\pi}  $mJcm$^{-2}/(50$~fs)$\approx$18.8~GWcm$^{-2}$.
The values shown in
Fig.~\ref{nonlinsurf_magy_vs_fermi_mimicopt_gamma25meV}
have been obtained for the smaller intensity of $I=$10~GWcm$^{-2}$,
for which we expect the corresponding smaller experimental peak current density
of 2.7mA/cm, which is larger than 0.78mA/cm by a factor of 3.5.
The laser-induced current density $J_y$ observed experimentally in Co/Pt
has been explained in terms of the IFE combined with the
inverse SOT~\cite{femtosecond_control_electric_currents_Huisman}.
Since 0.78mA/cm is only smaller by a factor of 3.5 compared to the
experimental value of 2.7mA/cm estimated for Co/Pt, we expect that this
\textit{magnetic circular photogalvanic effect} 
is in general a non-negligible contribution to $J_y$. If materials with small
IFE or small inverse SOT are used it is likely that the contribution from the
magnetic circular photogalvanic effect is dominant in $J_y$.

For circularly polarized light, 
the current density $J_x$ reaches 3A/m at $\mathcal{E}_{\rm F}=$6.8eV in 
Fig.~\ref{nonlinsurf_magy_vs_fermi_mimicopt_gamma25meV},
which is considerably larger than $J_y$. Also in the experiments 
on Co/Pt $J_x$ is found to be much larger 
than $J_y$~\cite{femtosecond_control_electric_currents_Huisman}.
In the experiments, $J_y$ depends strongly on the Pt thickness
and varies between 3.3A/m (1.3nm thick Pt) and
14.4A/m (3.9nm thick Pt) when a 50~fs laser pulse with fluence
1mJcm$^{-2}$ is used. We estimate that the corresponding 
current densities expected for the smaller intensity of $I=$10~GWcm$^{-2}$
range between 1.8A/m and 7.7A/m.
The experimentally measured $J_x$ in magnetic bilayer systems
has been interpreted 
to originate from the superdiffusive spin-current that is
converted into a charge current by the inverse spin Hall 
effect~\cite{thz_spin_current_kampfrath}.
This interpretation is supported by the 
very good correlation between the spin Hall conductivity
of the normal metal (NM) layer and
the measured THz amplitude
in Co$_{20}$Fe$_{60}$B$_{20}$(3nm)/NM(3nm) 
stacks~\cite{thz_emitter_Seifert}. 
Even though our theoretical values of $J_x$ shown in
Fig.~\ref{nonlinsurf_magy_vs_fermi_mimicopt_gamma25meV}
describe a magnetic photogalvanic effect and do not contain
the mechanism of generating a charge current by conversion of
a superdiffusive spin current, the maximal value of 3A/m
in Fig.~\ref{nonlinsurf_magy_vs_fermi_mimicopt_gamma25meV}
is non-negligible compared to the current density $J_x$ measured in
Co/Pt bilayer systems. Therefore, we expect that for suitable
material combinations the magnetic photogalvanic effect can
compete with the conversion of superdiffusive spin current
by the inverse spin Hall effect. In particular when the spin Hall
conductivity of NM is small or when the NM thickness is
much smaller than the hot-electron relaxation length~\cite{thz_emitter_Seifert}
we expect significant contributions from the 
magnetic photogalvanic effect to the current density $J_x$.

It might be possible to identify the contribution of the magnetic photogalvanic effect to the
current density $J_x$  in experiments by measuring the dependence of $J_x$ on the
polarization of light: According to 
Fig.~\ref{nonlinsurf_magy_vs_fermi_mimicopt_gamma25meV}
the current density $J_x$ generated by linearly polarized light depends strongly on whether the
light polarization vector is along $x$ or along $y$.
On the other hand, the generation of superdiffusive spin currents is not expected to
depend on the direction of the light polarization vector. Therefore, a strong
dependence of $J_x$ on the light polarization vector is a clear indication of the
magnetic photogalvanic effect.

Like the field-like contribution to the 
SOT~\cite{symmetry_spin_orbit_torques} the magnetic photogalvanic effect
is sensitive to the interfacial SOI in magnetic bilayer systems.
Therefore, we expect that the magnitude of the magnetic photogalvanic effect
is correlated with the magnitude of the field-like component of the SOT.
On the other hand, the contribution from the conversion of superdiffusive spin
current by the inverse spin Hall effect is expected to correlate with the
antidamping component of the SOT.
These two contributions to $J_x$ may therefore also be distinguished
in experiments via 
their different dependence on the interfacial SOI.
In contrast, we expect that it is more difficult to identify the contribution of
the magnetic photogalvanic effect to the current density $J_y$ in experiments,
because it competes with the current generated by the combined action of
the IFE and the inverse field-like SOT, i.e., both contributions to $J_y$ are
sensitive to the interfacial SOI.

\begin{figure}
\includegraphics[width=\linewidth]{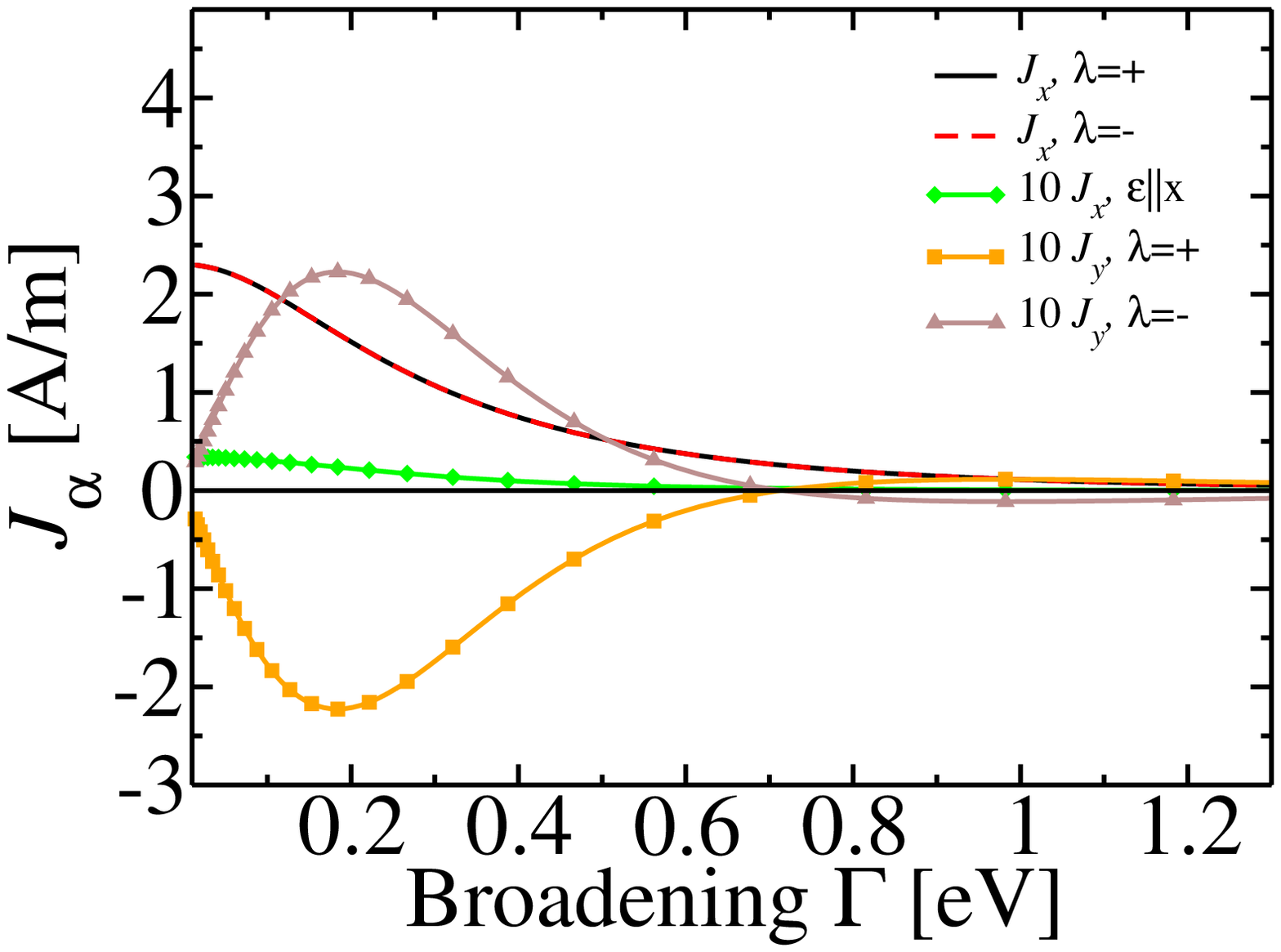}
\caption{\label{nonlin_magy_vs_sigma_alphacopt}
Laser-induced charge-current density $J_{\alpha}$
vs.\ broadening $\Gamma$
for $\hat{\vn{n}}$
in $y$ direction, $\alpha^{\rm R}=0.1$eV\AA\,, $\Delta V=1$~eV,
and $\mathcal{E}_{\rm F}=1.36$eV. 
Some curves have been multiplied by 
the factor 10 for better visibility, as indicated in the legend.
}
\end{figure}

Next, we discuss the dependence of the magnetic photogalvanic effect on 
the lifetime broadening $\Gamma$.
The laser-induced charge-current density is shown as a function
of $\Gamma$ in
Fig.~\ref{nonlin_magy_vs_sigma_alphacopt},
where we set $\mathcal{E}_{\rm F}=1.36$eV,
$\alpha^{\rm R}=0.1$eV\AA,
$\Delta V=1$~eV and $\hat{\vn{n}}$ points in $y$ direction.
While the current density $J_x$ decreases monotonically with increasing
broadening $\Gamma$ the current density $J_y$ exhibits a maximum at 
around $\Gamma=200$meV. This suggests that $J_y$ can be maximized
in magnetic bilayer systems such as Co/Pt by optimizing the interface roughness.
Indeed, recent experiments show that interface roughness is crucial
for the helicity-dependent photocurrent~\cite{PhysRevMaterials.3.084415}.

\begin{figure}
\includegraphics[width=\linewidth]{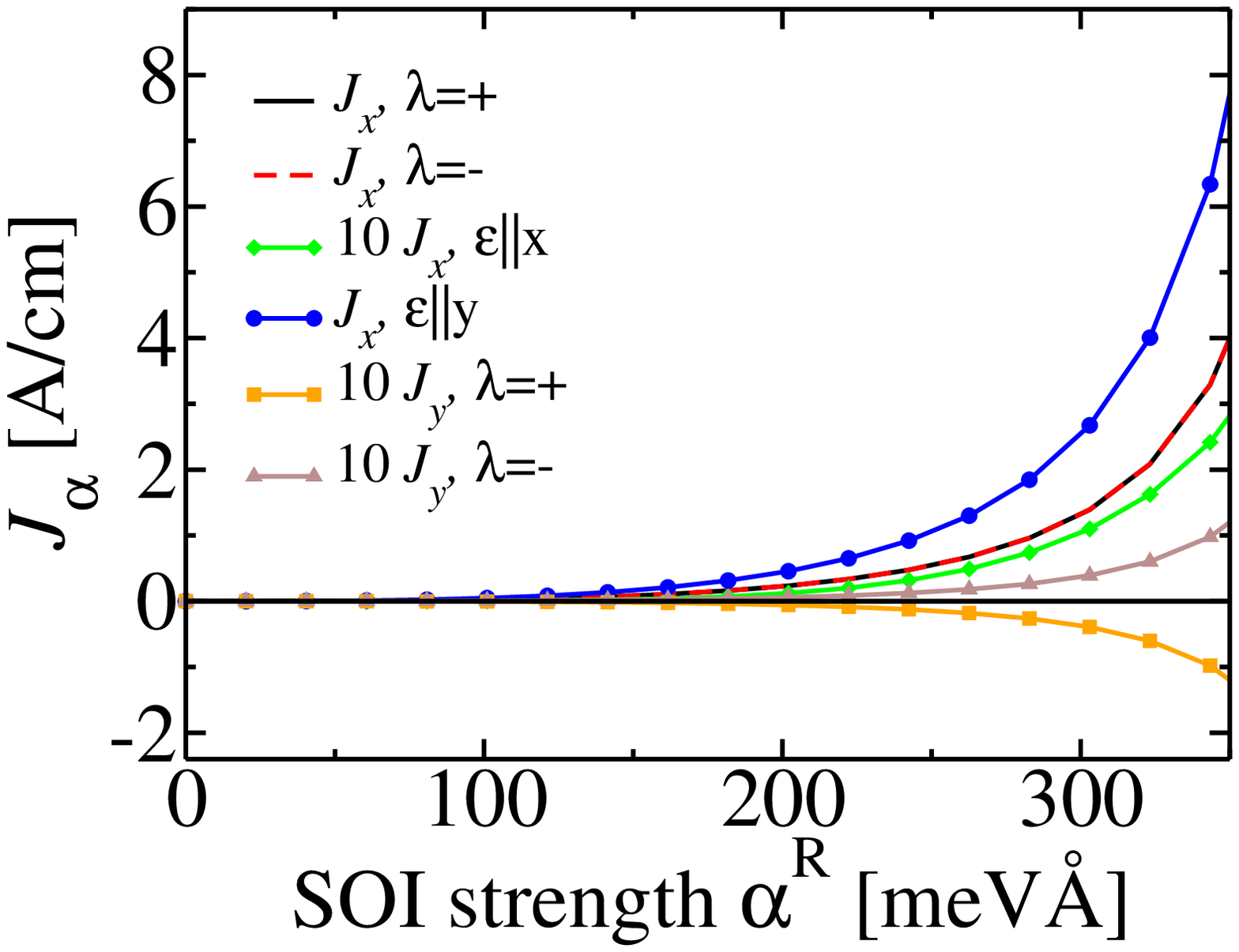}
\caption{\label{nonlin_magy_vs_alpha_gamma25meV}
Laser-induced charge-current density $J_{\alpha}$
vs.\ SOI strength $\alpha^{\rm R}$
for $\hat{\vn{n}}$
in $y$ direction, $\Delta V=1$~eV,
$\mathcal{E}_{\rm F}=1.36$eV, and $\Gamma=25$~meV. 
Some curves have been multiplied by 
the factor 10 for better visibility, as indicated in the legend.
}
\end{figure}

Finally, we discuss the dependence of the magnetic photogalvanic effect on 
the SOI-strength $\alpha^{\rm R}$.
The laser-induced charge-current density is shown as a function
of $\alpha^{\rm R}$ in
Fig.~\ref{nonlin_magy_vs_alpha_gamma25meV},
where we set $\mathcal{E}_{\rm F}=1.36$eV,
$\Delta V=1$~eV, $\Gamma=25$~meV 
and $\hat{\vn{n}}$ points in $y$ direction.
The magnetic photogalvanic effect increases strongly 
with the SOI strength $\alpha^{\rm R}$. 
Already for $\alpha^{\rm R}=200$meV\AA, which is
only twice as large as the value of $\alpha^{\rm R}$ used in 
Fig.~\ref{nonlinsurf_magy_vs_fermi_mimicopt_gamma25meV}
and in Fig.~\ref{nonlin_magy_vs_sigma_alphacopt},
the current density $J_{x}$ reaches 45.7 A/m
for linearly polarized light with polarization vector
along $y$ direction
and the current density $J_{y}$ reaches 0.59 A/m for circularly polarized light,
i.e., both $J_{x}$ and $J_{y}$ are larger than the maximal values
in Fig.~\ref{nonlinsurf_magy_vs_fermi_mimicopt_gamma25meV} by
one order of magnitude.
The laser-induced currents increase by another order of magnitude
when $\alpha^{\rm R}$ is increased further to 340meV\AA, 
where $J_{x}=630$~A/m
and $J_{y}=9.8$~A/m.

The question therefore arises whether $\alpha^{\rm R}$ can reach
300meV\AA\, or more in magnetic bilayer systems, in which
case the magnetic photogalvanic effect would be very strong
and would allow us to increase the efficiency of table-top 
THz emitters~\cite{thz_emitter_Seifert}
much further. 
While large Rashba SOI-strengths of $\alpha^{\rm R}=3.05$eV\AA\,
have been found in 
Bi/Ag(111) surface alloys~\cite{giant_spin_splitting_surface_alloying},
the estimated values of $\alpha^{\rm R}$ in magnetic
bilayer systems are often smaller, e.g.\ $\alpha^{\rm R}=0.095$eV\AA\,
in Co/Pt magnetic bilayer systems~\cite{sot_dmi_stiles}.
While one may expect that
magnetic bilayer systems with larger $\alpha^{\rm R}$ will be
discovered, it is likely that magnetic adatoms on surfaces
or 2D materials
are also systems with large magnetic photogalvanic effect.
For example in graphene decorated with W 
and in semi-hydrogenated
Bi(111) bilayers very large SOTs have been found in \textit{ab-initio}
calculations~\cite{mixed_weyl_sot}. 

\subsection{Laser-induced spin currents}
\label{sec_results_spin_current}
\begin{figure}
\includegraphics[width=\linewidth]{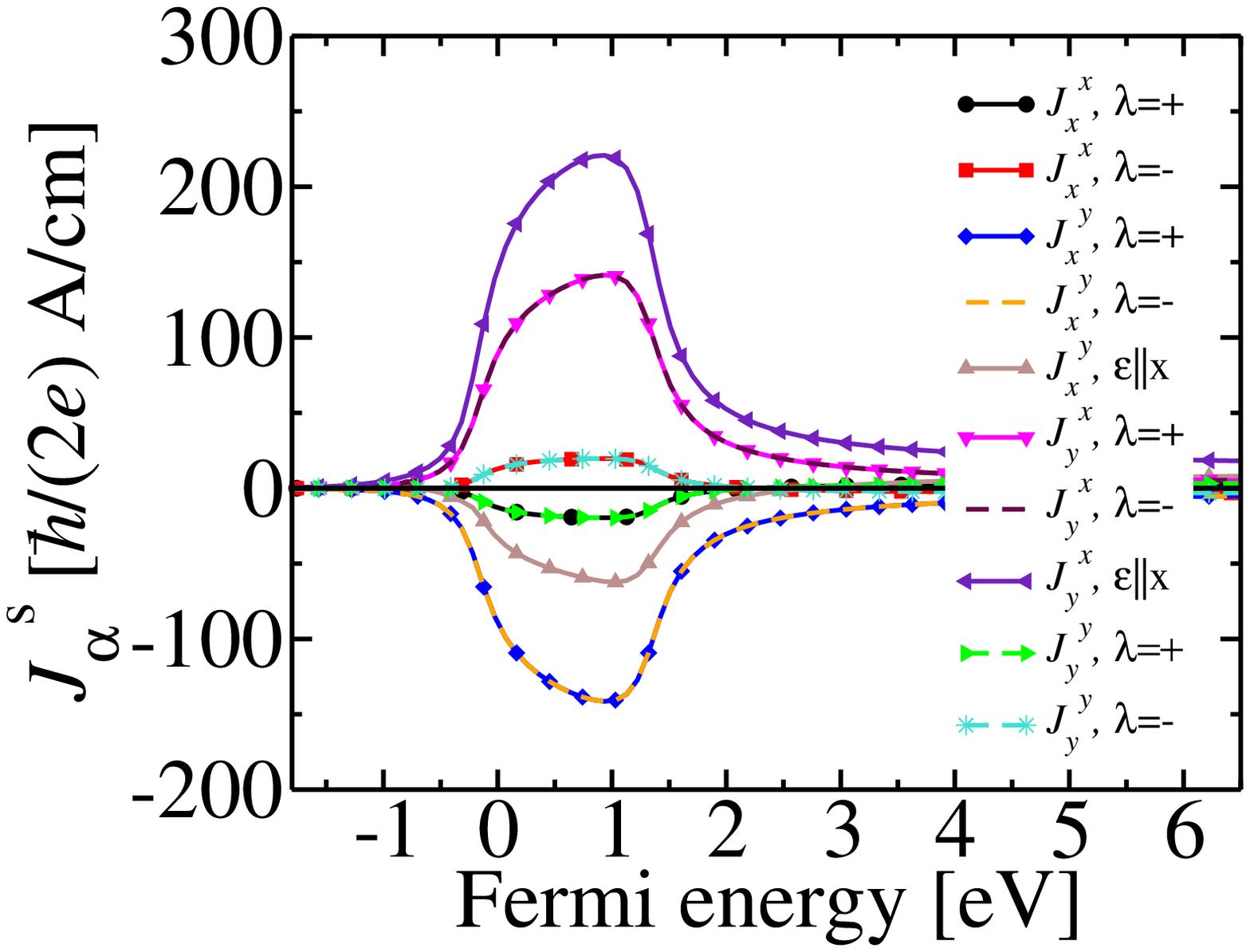}
\caption{\label{nolispicu_nonmagnetic_vs_fermi}
Laser-induced spin-current density $J_{\alpha}^{s}$
vs.\ Fermi energy
in the nonmagnetic Rashba model for the 
parameters $\alpha^{\rm R}=2$eV\AA\, and $\Gamma=136$meV.
}
\end{figure}

Next, we discuss the laser-induced spin currents in the nonmagnetic 
Rashba model~\cite{doi:10.1063/1.1882747}.
In Fig.~\ref{nolispicu_nonmagnetic_vs_fermi}
we show the laser-induced spin current as a function
of Fermi energy $\mathcal{E}_{\rm F}$
for the parameters $\alpha^{\rm R}=2$eV\AA\, and $\Gamma=136$meV. 
Since one electron carries a spin angular momentum of $\hbar/2$
it is convenient to discuss spin currents in units of $\hbar/(2e)$
times ampere. Therefore, we use in 
Fig.~\ref{nolispicu_nonmagnetic_vs_fermi} $\hbar/(2e)$A/cm as unit
of the spin current density.
A spin current of 1~A~$\hbar/(2e)$ can be thought
of as a positive charge current of 0.5~A carried by spin-down electrons
accompanied by a negative charge current of 0.5~A carried by spin-up electrons. 
In agreement with the discussion in section~\ref{sec_symmetry_spicu} 
summarized in Table~\ref{tab_lasincuspi_magy}
the following components are nonzero in the nonmagnetic
case: For linearly polarized light only $J_{y}^{x}$ and $J_{x}^{y}$
are allowed by symmetry.
For circularly polarized light $J_{y}^{x}$ and $J_{x}^{y}$ 
(even in the helicity $\lambda$)
and $J_{y}^{y}$ and $J_{x}^{x}$ 
(odd in the helicity $\lambda$) are allowed by symmetry.
Light linearly polarized along $x$ induces a smaller $J_{x}^{y}$ than
circularly polarized light. However, light linearly polarized along $x$
induces a larger $J_{y}^{x}$ than circularly polarized light.
For circularly polarized light rotation around the $z$ axis by $90^{\circ}$
is a symmetry operation, which leads to $J_{x}^{y}=-J_{y}^{x}$ 
and to $J_{x}^{x}=J_{y}^{y}$.

For the parameter range covered by Fig.~\ref{nolispicu_nonmagnetic_vs_fermi}
the helicity-odd effects are much smaller than the helicity-even effects.
The maximum spin-current density in Fig.~\ref{nolispicu_nonmagnetic_vs_fermi} is
attained by the component $J_{y}^{x}$ for light polarized linearly 
along $x$ and it amounts to $J_{y}^{x}=221\hbar/(2e)$~A/cm. 
This spin-current density can be thought of as a charge-current 
density of spin-up electrons 
(spin-up and spin-down refer to the $x$ axis as spin-quantization axis) of
110~A/cm flowing into the negative $y$ direction and a 
charge-current density of
spin-down electrons of 110~A/cm flowing into the positive $y$ direction.
This spin-dependent charge-current density of $\pm 110$~A/cm 
exceeds the laser-induced charge-current density
that has been measured experimentally in
magnetic bilayer
systems~\cite{thz_spin_current_kampfrath,thz_emitter_Seifert,femtosecond_control_electric_currents_Huisman}
at comparable light intensity by several orders of magnitude. 
Since the net charge current is zero it does not generate a THz
electromagnetic signal, which makes this effect difficult to
observe experimentally. The inverse spin Hall effect could be used
to convert these spin currents into detectable charge currents.
However, this would
require to inject the spin current from the Rashba system
into a different system, because in the
Rashba model itself there is no inverse spin Hall effect that converts
any of the spin-current densities $J_{y}^{x}$, $J_{x}^{y}$, $J_{y}^{y}$ or $J_{x}^{x}$ into
a charge current.

\begin{figure}
\includegraphics[width=\linewidth]{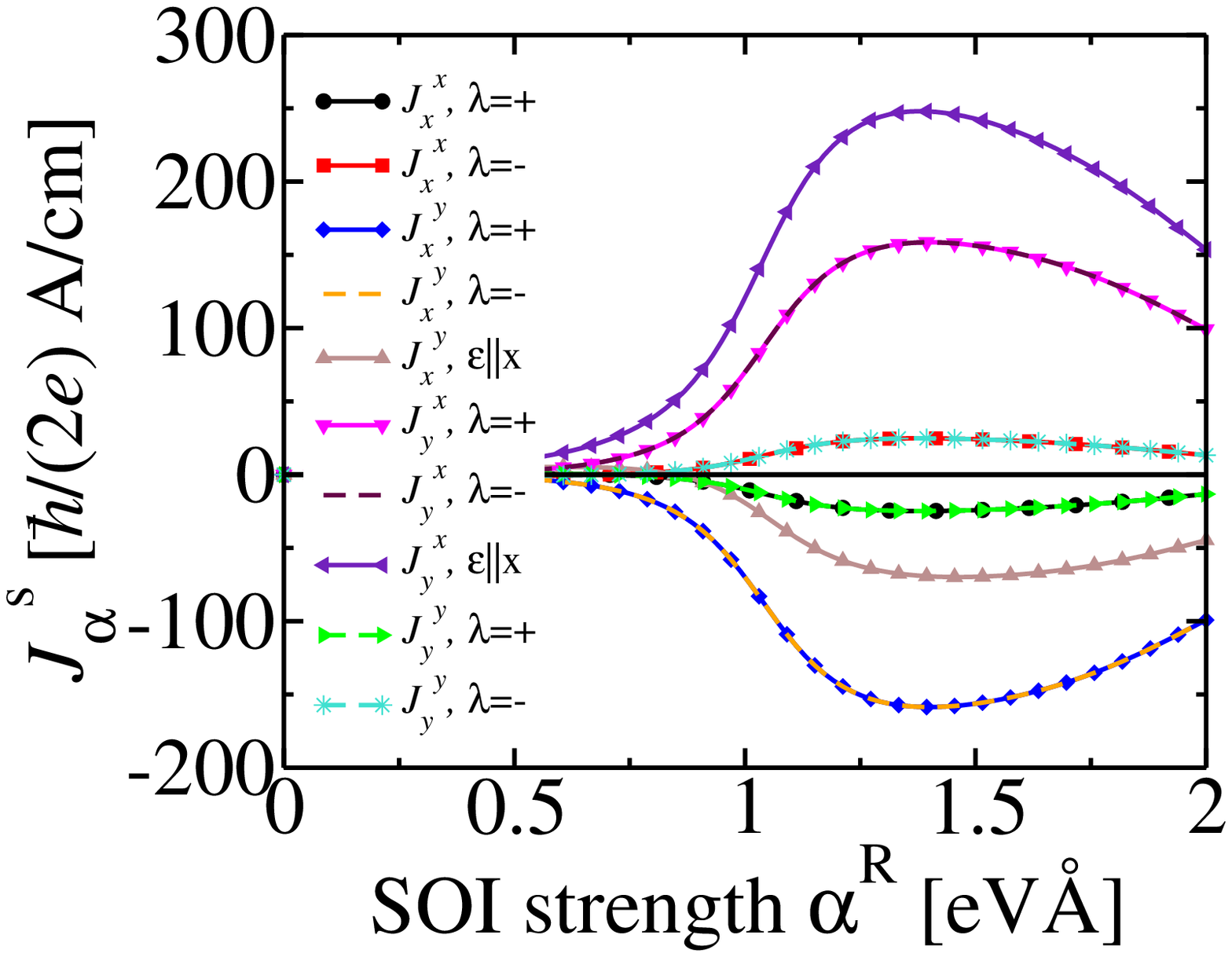}
\caption{\label{nolispicu_nonmag_vs_alpha}
Laser-induced spin-current density $J_{\alpha}^{s}$
vs.\ SOI-strength $\alpha^{\rm R}$
in the nonmagnetic Rashba model
for the parameters $\mathcal{E}_{\rm F}=1.36$eV
and $\Gamma=136$meV.
}
\end{figure}

\begin{figure}
\includegraphics[width=\linewidth]{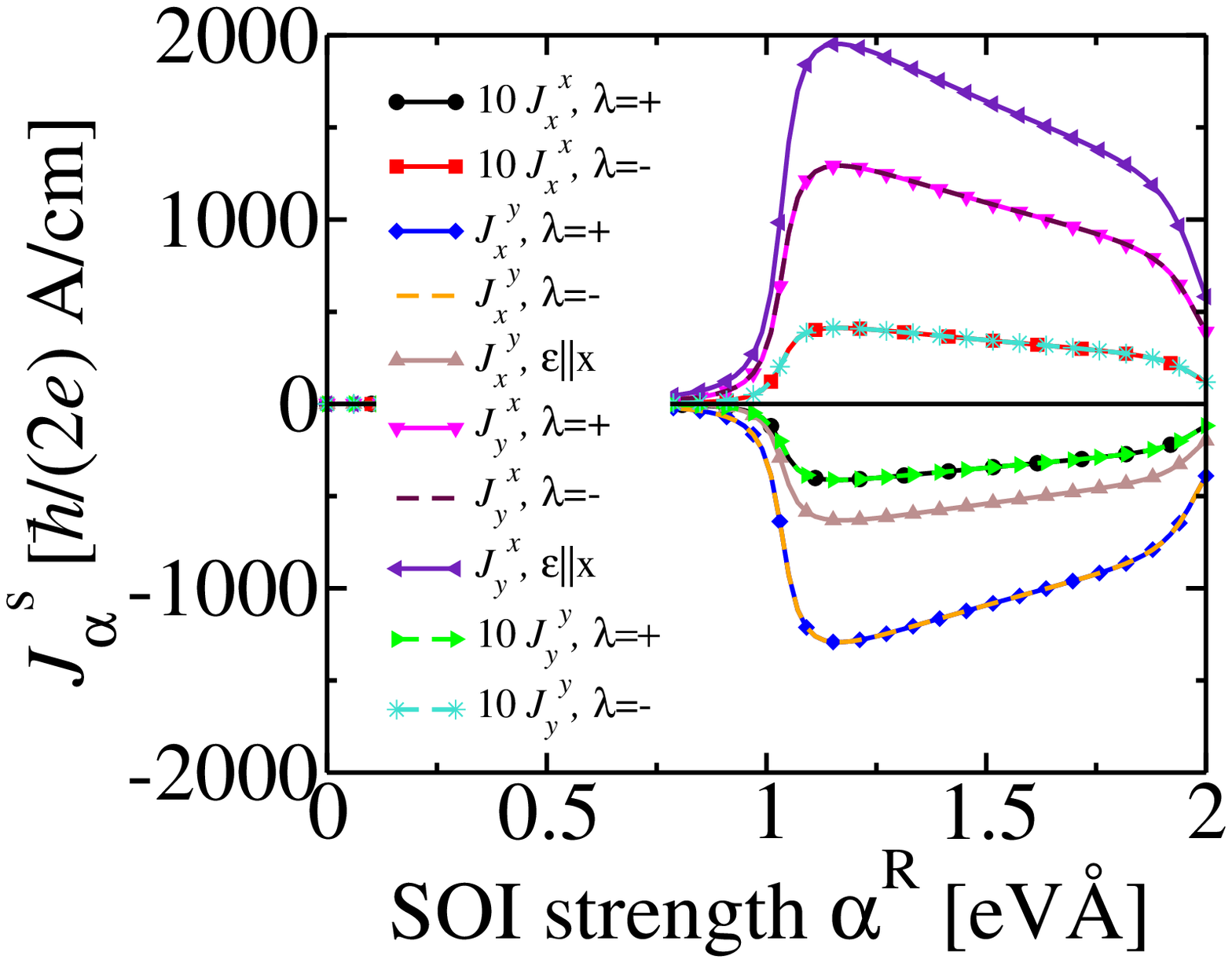}
\caption{\label{nolispicu_vs_alpha_gamma25meV}
Laser-induced spin-current density $J_{\alpha}^{s}$
vs.\ SOI-strength $\alpha^{\rm R}$
in the nonmagnetic Rashba model
for the parameters $\mathcal{E}_{\rm F}=1.36$eV
and $\Gamma=25$~meV.
Some curves have been multiplied by 
the factor 10 for better visibility, as indicated in the legend.
}
\end{figure}

In Fig.~\ref{nolispicu_nonmag_vs_alpha}
we show the laser-induced spin-current density as a function
of SOI strength $\alpha^{\rm R}$
for the parameters $\mathcal{E}_{\rm F}=1.36$eV 
and $\Gamma=136$meV. The figure shows that for
SOI strengths larger than $\alpha^{\rm R}=1$eV\AA\, the effect is
particularly sizable. In Fig.~\ref{nolispicu_vs_alpha_gamma25meV}
we show the laser-induced spin-current density as a function
of SOI strength $\alpha^{\rm R}$ for the smaller
broadening of $\Gamma=25$meV at the 
Fermi energy $\mathcal{E}_{\rm F}=1.36$eV.
For this smaller broadening much larger spin-current densities can be reached. 

\section{Summary}
\label{sec_summary}
We study the laser-induced charge current in the ferromagnetic
Rashba model with in-plane magnetization and predict that this
\textit{magnetic photogalvanic effect} 
is sufficiently strong in magnetic bilayer systems to be observable in 
experiments. The magnetic photogalvanic effect has one 
component that is odd in the helicity of light and a second
component that is even in the helicity of light. The
helicity-odd component can be maximized by optimizing the amount of
disorder in the system. The helicity-even component depends
strongly on the direction of the light-polarization vector when
linearly polarized light is used. 
Additionally, we discuss laser-induced spin currents
in the nonmagnetic Rashba model. Thereby, we predict
that laser-induced pure spin currents at nonmagnetic surfaces
and interfaces with giant Rashba effect
exceed the laser-induced charge currents in magnetic bilayer
systems such as Co/Pt by several orders of magnitude.

\section*{Acknowledgments}We acknowledge financial support from
Leibniz Collaborative Excellence project OptiSPIN $-$ Optical Control
of Nanoscale Spin Textures, and funding  under SPP 2137 ``Skyrmionics"
of the DFG. 
We gratefully acknowledge financial support from the European Research
Council (ERC) under 
the European Union's Horizon 2020 research and innovation 
program (Grant No. 856538, project ``3D MAGiC''), and ITN Network
COMRAD. 
The work was also supported by the Deutsche Forschungsgemeinschaft 
(DFG, German Research Foundation) $-$ TRR 173 $-$ 268565370 (project
A11), TRR 288 $-$ 422213477 (project B06).  We  also gratefully
acknowledge the J\"ulich 
Supercomputing Centre and RWTH Aachen University for providing
computational 
resources under project No. jiff40.

\bibliography{lasincucspira}

\end{document}